\documentclass[lettersize,journal]{IEEEtran}
\usepackage{amsmath,amsfonts}
\usepackage{algorithmic}
\usepackage{algorithm}
\usepackage{array}
\usepackage[caption=false,font=normalsize,labelfont=sf,textfont=sf]{subfig}
\usepackage{textcomp}
\usepackage{stfloats}
\usepackage{url}
\usepackage{verbatim}
\usepackage{graphicx}
\usepackage{cite}
\usepackage[colorlinks,linkcolor=red]{hyperref}
\usepackage{booktabs}
\usepackage{multirow}

\begin{document}

\title{CWT-Net: Super-resolution of Histopathology Images Using a Cross-scale Wavelet-based Transformer}
\author{Feiyang Jia, Zhineng Chen, Ziying Song, Lin Liu and Caiyan Jia.
\thanks{This work is supported by the National Natural Science Foundation of China under Grant No. 62172103.\emph{(Corresponding author: Caiyan Jia.)}}
\thanks{Feiyang Jia, Ziying Song, Lin Liu, Caiyan Jia are with School of Computer and Information Technology, Beijing Jiaotong University, Beijing 100044, China (e-mail: feiyangjia@bjtu.edu.cn; songziying@bjtu.edu.cn; 23120379@bjtu.edu.cn; cyjia@bjtu.edu.cn.)}
\thanks{Zhineng Chen is with School of Computer Science, Fudan University, Shanghai 200433, China (e-mail: zhinchen@fudan.edu.cn.)}
}




\maketitle

\begin{abstract}
Super-resolution (SR) aims to enhance the quality of low-resolution images and has been widely applied in medical imaging. We found that the design principles of most existing methods are influenced by SR tasks based on real-world images and do not take into account the significance of the multi-level structure in pathological images, even if they can achieve respectable objective metric evaluations. In this work, we delve into two super-resolution working paradigms and propose a novel network called CWT-Net, which leverages cross-scale image wavelet transform and Transformer architecture. Our network consists of two branches: one dedicated to learning super-resolution and the other to high-frequency wavelet features. To generate high-resolution histopathology images, the Transformer module shares and fuses features from both branches at various stages. Notably, we have designed a specialized wavelet reconstruction module to effectively enhance the wavelet domain features and enable the network to operate in different modes, allowing for the introduction of additional relevant information from cross-scale images. Our experimental results demonstrate that our model significantly outperforms state-of-the-art methods in both performance and visualization evaluations and can substantially boost the accuracy of image diagnostic networks.
\end{abstract}



\section{Introduction}\label{sec1}

\begin{figure}[htbp]%
\centering
\includegraphics[width=0.48\textwidth]{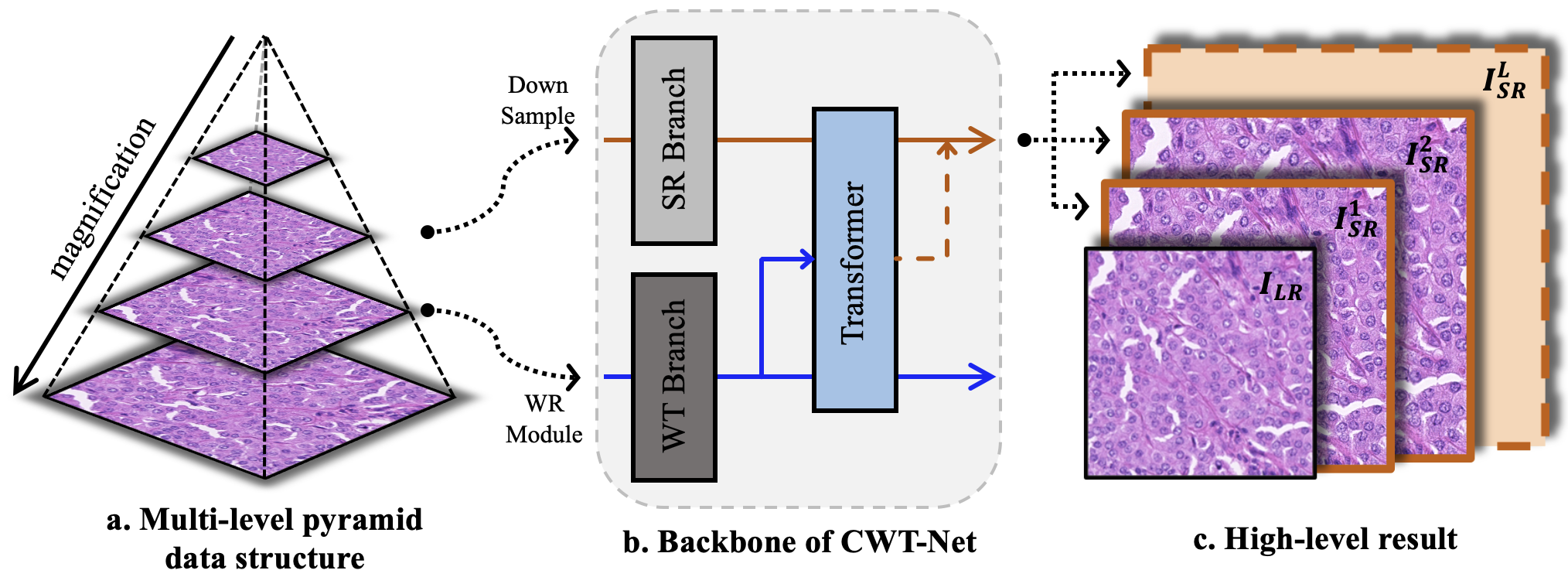}
\caption{a. Digital medical images typically have multiple levels and dimensions, requiring higher generation and storage costs. Medical images with a pyramid structure can store images at multiple magnifications, facilitating healthcare professionals in their examination. However, existing SR methods have not taken the above-mentioned characteristics into account. b. To tackle this issue, we propose CWT-Net, which combines a multitasking strategy to expand the utilization of pyramid shaped data. c. CWT-Net aims to generate high-level digital medical images at a lower cost.}\label{fig1}
\end{figure}

Super-resolution (SR) is a pivotal task within the realm of computer vision, focusing on the enhancement of high-resolution (HR) images using information extracted from their low-resolution (LR) counterparts \cite{bib1}. Interpolation algorithms are commonly employed for SR task, include nearest neighbor interpolation, bilinear interpolation, bicubic interpolation, and more \cite{bib1}. Traditional Single Image Super-resolution (SISR) techniques often rely on substantial prior information to achieve image reconstruction \cite{bib2,bib3}, and these methods emphasize information correlation across different frequency domains but tend to overlook the specific degradation process associated with low-quality images. In recent years, significant advancements have been made in the field of SR tasks, primarily focusing on real-world images \cite{bib4,bib5,bib6,bib8,bib9,bib10,bib11,bib12,bib13,bib31,bib44,bib45,bib47,bib48,bib49,bib50,bib51,bib52,bib53}. 


Medical images play a crucial role in clinical diagnosis, treatment planning, and quantitative analysis of human organs and tissues. However, acquiring high-resolution medical images is often a time-consuming and costly process. SR methods offer a cost-effective solution for enhancing the quality of medical images. Zhao X {\it et al.} \cite{bib20} introduced CSN, a model designed to enhance the quality of magnetic resonance images (MRI). Qiao C {\it et al.} \cite{bib21} created the BioSR dataset and developed the DFCAN/DFGAN for scientific microscope images. Li Z {\it et al.} \cite{bib22} proposed TSMLSRNet, a method aimed at reconstructing arterial spin-labeled perfusion MRI images. L Mukherjee {\it et al.} \cite{bib23} utilized medium-level resolution images from Whole Slide Images (WSI) to reconstruct LR images. Chen Z {\it et al.} \cite{bib27} developed SWD-Net, a model focused on reconstructing breast cancer histopathology images using wavelet domain features. Wu X {\it et al.} \cite{bib42} presented MMSRNet that tackles multiple SR tasks with different magnifications as a unified, joint task. Wang H {\it et al.} \cite{bib54} proposed $W^{2}$AMSN for MR images, this network extracts features of different sizes at multiple scales and designs a non-reduction attention mechanism to recalibrate feature responses adaptively. Qiu Z {\it et al.} \cite{bib46} proposed DS2F, which uses super-resolution tasks as an auxiliary to medical image segmentation tasks.

The existing studies, despite achieving impressive results in objective metrics, are still constrained by their working paradigms. Image SR research typically operates within two paradigms: SISR \cite{bib4,bib5,bib6,bib9,bib10,bib11,bib12,bib16,bib17,bib20,bib21,bib22,bib24,bib27,bib28,bib29,bib30,bib31} and RefSR (Reference-Based Super-Resolution) \cite{bib15,bib32,bib33,bib34}. In the SISR paradigm, network models can rapidly learn the LR data distribution. Increasing network depth generally improves model performance, but this often comes at the cost of increased blurring and artifacts in the reconstructed images. The RefSR paradigm enhances the details of reconstructed images by gathering supplementary information from reference images. The stronger the similarity between the reference image, the LR image, and the HR image, the better the SR network's reconstruction performance. However, the RefSR approach doesn't effectively learn how LR images degrade, which is a limitation. Furthermore, most SR methods applied to medical images do not adequately account for the continuous sequence formats of medical images. For instance, WSI employ a pyramid-like structure (see Fig.\ref{fig1}.a) to store data at multiple sampling levels. If SR methods could mine and incorporate continuous structural information, it would lead to superior reconstruction results.

Transformers \cite{bib14} have gained significant traction in computer vision tasks, offering multitasking capabilities \cite{bib28} and effective information sharing among different domains \cite{bib15}. Meanwhile, wavelet transforms have proven valuable for decomposing images into low-frequency and high-frequency components \cite{bib13,bib31}. Inspired by the powerful performance of Transformers and recognizing the multi-scale nature of medical images, especially in pathology, we introduce CWT-Net. This network leverages multi-scale pathology images as input and consists of two branches: the Super-resolution Branch (SR Branch) processes LR images by feature extraction and upsampling to produce high-quality SR results, while the Wavelet Transform Branch (WT Branch) extracts high-frequency details from HR images across various scales using wavelet transforms (see Fig.\ref{fig1}.b). The Transformer module progressively merges information from both branches, enhancing the primary functions of the SR branch to get the high-level result (see Fig.\ref{fig1}.c).

Our contributions are as follows:

(1) We introduce an end-to-end model, CWT-Net, which efficiently captures high-frequency details in pathology images across various scales, expediting the learning process for SR tasks. CWT-Net outperforms existing pathology image reconstruction and SR models, representing a significant advancement.

(2) We devise a specialized wavelet reconstruction module to enhance wavelet information at a single scale. This module allows for the utilization of additional cross-scale information while adhering to the SISR working paradigm. Consequently, the training and testing phases of CWT-Net adopt different strategies. When CWT-Net is employed as a pre-training model for other SR networks, this module endows the SR network with the advantages of both working paradigms, eliminating the need for extensive information construction.

(3) We curate the benchmark dataset MLCamSR, consisting of sampling regions with three levels of real sampled images. This dataset enables CWT-Net to be trained with undegraded cross-scale information, further enhancing its performance.

\section{Related Work}\label{sec2}

\subsection{Working paradigm of SR task}\label{subsec2.1}

Most deep learning-based SR methods operate under the SISR paradigm. Their fundamental approach revolves around designing multiple pathways to extract valuable information from the limited data available in LR images, including residual structures \cite{bib6,bib9,bib10,bib11,bib30}, generative structures \cite{bib6,bib12,bib24,bib51}, attention mechanisms \cite{bib10,bib16,bib28,bib43,bib53}, and digital image spectrum analysis \cite{bib13,bib27,bib28,bib31,bib48,bib49}. For instance, Chen Z {\it et al.} \cite{bib27} proposed SWD-Net for the SR task of breast cancer histopathology images. In this work, the wavelet-aware convolution (WAC) module can effectively extract the context information in the wavelet domain, while the wavelet feature adaptation (WFA) module adjusts the wavelet coefficients to an appropriate range. Shahidi \cite{bib24} introduced WA-SRGAN with Generate Adversarial Networks structure \cite{bib7} to reconstruct breast cancer histopathology images, evaluating the model using various metrics like PSNR \cite{bib25,bib26}, SSIM \cite{bib26}, and MSE.

RefSR methods \cite{bib15,bib32,bib33,bib34} aim to reconstruct HR images using additional information from reference images, requiring techniques to identify and align similar regions between the reference and LR images. TTSR \cite{bib15} utilizes Transformer to jointly learn LR and reference image relationships for accurate texture transfer.

However, RefSR methods are inherently constrained by the availability and quality of reference images \cite{bib15,bib34}. CWT-Net leverages the hierarchical structure of histopathology images to introduce cross-scale wavelet information while maintaining the SISR paradigm. Specifically, we design the wavelet reconstruction module to approximate cross-scale information from LR images, replacing the WT branch's information flow during network testing. This strategy enables CWT-Net to harness multi-level information during training and overcome cross-scale limitations during testing for more effective histopathology image super-resolution.

\subsection{Transformer in SR Task}\label{subsec2.2}

The Transformer architecture \cite{bib14} has been swiftly integrated into SR tasks \cite{bib15,bib16,bib17,bib18,bib19} since its inception.
Yang F {\it et al.} \cite{bib15} introduced TTSR, a method utilizing the Texture Transformer to transfer texture details from high-resolution reference images to LR reconstructions. In TTSR, features from both the reference and LR images are extracted using the Learnable Texture Extractor (LTE). The texture features, denoted as Q, K, and V, are derived from the upsampled LR image, the reference image subjected to downsampling and subsequent upsampling, and the original reference image, respectively. Following this, Hard-Attention (HA) computes and transfers similar features from the reference image, while Soft-Attention (SA) combines these similar features through weighted fusion. Feng C {\it et al.} \cite{bib28} proposed T2Net, a framework that combines MRI reconstruction and SR tasks for multi-task learning by Task Transformer module.


Compared to CNN-based methods, the Transformer's self-attention mechanism effectively captures global image features, enhancing the model's expressiveness. In a similar vein, CWT-Net's objective is to establish relationships between cross-scale wavelet transform features and LR image features using Transformer, thus bolstering its performance in the SR task.

\section{Method}\label{sec3}

\subsection{Overall Structure}\label{subsec3.1}

\begin{figure*}[htbp]%
\centering
\includegraphics[width=1\textwidth]{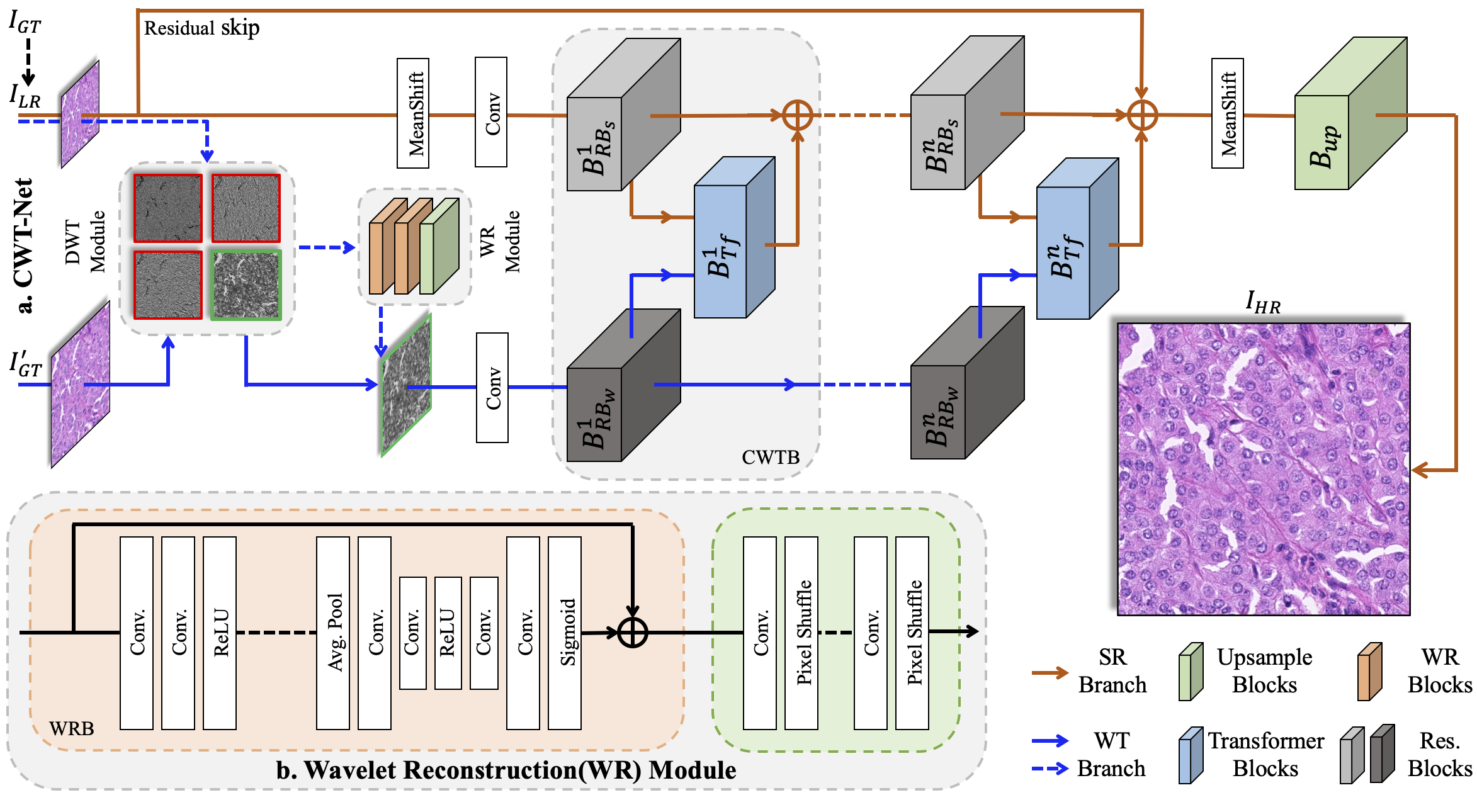}
\caption{a. CWT-Net's Comprehensive Architecture: CWT-Net adopts a dual-branch architecture, each dedicated to distinct tasks. The SR branch (orange) focuses on transforming LR images into HR images. The WT branch (blue) is designed to capture wavelet features within HR images across multiple scales. Transformer blocks play a pivotal role in identifying analogous textures among cross-level wavelet features, which are then amalgamated with the SR branch's output. b. Wavelet Reconstruction (WR) Module: The WR module, featured here, integrates a channel attention mechanism characterized by deeper and wider inner channels to facilitate feature extraction.}\label{fig2}
\end{figure*}

To effectively address the challenge of obtaining high-resolution images from multi-level histopathology data, we introduce a end-to-end framework named CWT-Net.

CWT-Net comprises three integral components: the SR branch, the WT branch, and the Transformer module. In the initial stages of processing, histopathology images $I_{GT}$ from the training set undergo a downsampling procedure using an interpolation-based degradation algorithm. This results in the creation of corresponding LR images, denoted as $I_{LR}$. To make the most of the intermediate structural information within the histopathology images, we execute an additional round of downsampling for each $I_{GT}$ to generate $I_{GT}^{'}$. Subsequently, $I_{LR}$ serves as the input to the SR branch, while $I_{GT}^{'}$ acts as the input to the WT branch. Importantly, $I_{GT}$ and $I_{GT}^{'}$ share identical sampling center coordinates, and their sampling radii are inversely proportional to the data level. When $I_{GT}^{'}$ is scaled to match the resolution of $I_{GT}$, both images encompass an equivalent range of histopathology sections. It's vital to ensure that $I_{GT}^{'}$ adheres to the condition expressed in Eq. (1):

\begin{equation}
Lv_{GT} \ge Lv_{gt^{'}} \ge Lv_{LR}
\end{equation}
where $Lv_{GT}$, $Lv_{gt^{'}}$, and $Lv_{LR}$ denote the sampling levels of $I_{GT}$, $I_{GT}^{'}$, and $I_{LR}$, respectively. When $Lv_{gt^{'}}=Lv_{GT}$, the wavelet information harnessed and provided by the WT branch wholly encompasses the wavelet information present in the LR image. Conversely, when $Lv_{gt^{'}}=Lv_{LR}$, CWT-Net essentially reverts to the SISR paradigm, with the features of the WT branch failing to contribute additional information. We do not consider scenarios where $Lv_{gt^{'}}$ exceeds $Lv_{GT}$ or falls below $Lv_{LR}$: the former renders the SR task nonsensical (an image with more information than $I_{GT}$ cannot be employed as a prior for the SR task), while the latter contradicts the expectations of the WT branch.

Subsequently, the data traverse through a series of Cross-scale Wavelet-based Transformer Blocks (CWTBs), each comprising a residual module for the SR branch, a residual module for the WT branch, and a Transformer block. Towards the conclusion of the SR branch, the features derived from $I_{LR}$ pass through the upsampling module $B_{up}$ to produce the SR outcome, denoted as $I_{HR}$.

\subsection{SR Branch}\label{subsec3.2}

The SR branch in CWT-Net serves as the primary component for executing its core task, as illustrated in Fig.\ref{fig2}.a. For an input image $I_{LR}$, regardless of its dimensions, we initially define the feature as $f_{sr}$. Within the SR branch, a MeanShift layer is employed to normalize the color channels by subtracting their means, followed by a convolution layer that produces the initial shallow feature $f_{sr}^{0}$. Subsequently, $f_{sr}^{0}$ proceeds through the SR branch's backbone to undergo further feature extraction. This backbone is comprised of several residual blocks denoted as $B_{RB_{s}}$. Each $B_{RB_{s}}$ contains multiple sets of residual structures.

Define CWT-Net containing n CWTBs, and denote $f_{sr}^{n-1}$ and $f_{hwt}^{n-1}$ as the features within both branches after the (m-1)-th CARBs ($m-1\leq n$). For the m-th CWTB, the feature flow process in the SR branch is described by Eq. (2):

\begin{equation}
\begin{split}
f_{s r}^{m}=B_{R B_{s}}^{m}\left(f_{s r}^{m-1}\right) 
\bigoplus B_{T f}^{m}\left\{B_{R B_{s}}^{m}\left(f_{s r}^{m-1}\right), B_{R B_{w}}^{m}\left(f_{h w t}^{m-1}\right)\right\}
\end{split}
\end{equation}
where $\bigoplus$ denotes element-wise summation, $f_{sr}^{m}$ is the feature of the SR branch after the m-th CWTB, $B_{RB_{s}}^{m} (.)$ and $B_{RB_{s}}^{w} (.)$ denote the m-th residual block of the SR branch and WT branch (Section \ref{subsec3.3}), respectively, and $B_{Tf}^{m} (.,.)$ is the m-th Transformer block (Section \ref{subsec3.4}).

The SR branch in CWT-Net incorporates a long skip connection, as we have observed that further stacking of CWTBs does not yield improved performance. The primary aim of this long skip connection is to convey shallow features $f_{sr}^{0}$ to the lower depths of the network, to compensate for the potential loss of low-frequency information. This process is represented by Eq. (3):

\begin{equation}
f_{s r}^{\prime}=f_{s r}^{0} \bigoplus f_{s r}^{n}
\end{equation}
where $f_{sr}^{'}$ represents the comprehensive feature generated by the SR branch's backbone network. The nested residual connections within the SR branch facilitate the utilization of maximum contextual information within the network's constrained depth. Upon reaching the end of the SR branch, $f_{sr}^{'}$ initially undergoes the MeanShift layer to restore the original channel means. Subsequently, it passes through the $B_{up}$, which incorporates the subpixel shuffle layer to incrementally scale the image to the desired magnification levels.

\subsection{WT Branch}\label{subsec3.3}

The WT branch's purpose is to extract wavelet transform features from the multi-scale image $I_{GT}^{'}$. We initialize the feature in the WT branch as $f_{wt}$. To capture wavelet transform features, $f_{wt}$ first undergoes a 2D discrete wavelet transform (DWT) module \cite{bib35} to produce the feature $f_{hwt}$. Specifically, the DWT utilizes the filter $F_{HH}$ for convolution with $f_{hwt}$ to obtain sub-features. The filter $F_{HH}$ is defined as shown in Eq. (4):

\begin{equation}
F_{H H}=\frac{1}{2}\left[\begin{array}{cc}1 & -1 \\-1 & 1\end{array}\right]
\end{equation} 

Following the DWT module, the feature $f_{hwt}$ encodes the diagonal structural details of $I_{GT}^{'}$. We opt for the Haar wavelet for implementing the DWT module. In histopathology images, dense high-frequency information often arises from tissue cell membrane edges. The Haar wavelets effectively describe the high-frequency characteristics of histopathology images \cite{bib12}. Notably, $f_{hwt}$ shares the same dimensions as $f_{sr}$ from the SR branch because the sub-features generated by wavelet transform at each level are only half the size of the original features.

Subsequently, the feature $f_{hwt}$ is passed through a convolutional layer, resulting in the initial feature $f_{hwt}^{0}$ for the WT branch. The backbone network of the WT branch is composed of several residual blocks $B_{RB_{w}}$. The feature flow within the backbone network of the WT branch is defined as shown in Eq. (5):

\begin{equation}
f_{h w t}^{m}=B_{R B_{w}}^{m}\left(f_{h w t}^{m-1}\right)
\end{equation}
where m, $B_{RB_{w}}^{m} (.)$, $f_{hwt}^{m-1}$, and $f_{hwt}^{m}$ are defined in the same manner as Eq. (2). It's important to note that the depth of $B_{RB_{w}}$ varies based on the upsampling factor in the SR task (Section \ref{subsec4.2}). The disparity in distance scaling among the four channels affects the wavelet transform differently. High-frequency channels with larger distance scaling disparities might lead the network to disregard wavelet coefficients with smaller values. In terms of texture features, the original features from the wavelet transform exhibit sharp but incoherent high-frequency edges. This becomes especially pronounced as the average distance between $I_{GT}^{'}$ and $I_{LR}$ increases, particularly in scenarios with higher upsampling factors. Hence, it's crucial to dynamically adjust the depth of the WT branch to address these issues.

To enable the WT branch to operate within the SISR paradigm, we have introduced the wavelet reconstruction (WR) block, as depicted in Fig. \ref{fig2}.b. In situations where $I_{GT}^{'}$ is unavailable to the WT branch, such as during testing or when $I_{GT}^{'}$ is not provided, we want the WT branch to utilize approximate features from $I_{LR}$ to complete its task. The WR block is constructed based on the channel attention (CA) mechanism. Specifically, the channel attention mechanism employs global average pooling to transform global spatial information into channel descriptors and uses the sigmoid activation function. The ReLU \cite{bib36} layer is then applied to further accentuate the weight differences between features. The WR block employs a smaller feature downsampling factor to extend the weighting range. Similar to the backbone network, the depth of the WR module varies based on the upsampling factor (Section \ref{subsec4.2}).

For the feature $f_{hwt}$ with a size of $C \times H/S \times W/S$, where S is a multiple of the SR task, the average pooling layer reduces $f_{hwt}$ to $1 \times 1$ to obtain the primary representation $f_{g}$ of the channel, as described in Eq. (6):

\begin{equation}
f_{g}=\operatorname{AvgPool}\left(f_{h w t}\right)=\frac{1}{\mathrm{H} \times \mathrm{W}} \sum_{i=0}^{H} \sum_{j=0}^{W} f_{h w t}(i, j)
\end{equation} 
where $f_{hwt}(i,j)$ denotes the value at position (i,j) in $f_{hwt}$, and $AvgPool(.)$ represents global average pooling. Subsequently, $f_{g}$ is described as Eq. (7):

\begin{equation}
f_{w r}^{m}=\partial\left(F_{\uparrow}\left(\delta\left(F_{\downarrow}\left(f_{g}\right)\right)\right)\right)
\end{equation} 
where $f_{wr}^{m}$ denotes the output of the m-th WR block, $\delta(.)$ and $\sigma(.)$ represent the ReLU and sigmoid activations, and $F_{\downarrow} (.)$ and $F_{\uparrow} (.)$ denote the feature downsampling convolution and feature upsampling convolution performed by CAR. Finally, $f_{wr}$ is upsampled and amplified to the size of $C \times H \times W$.

\subsection{Transformer Block}\label{subsec3.4}

\begin{figure*}[htbp]%
\centering
\includegraphics[width=0.8\textwidth]{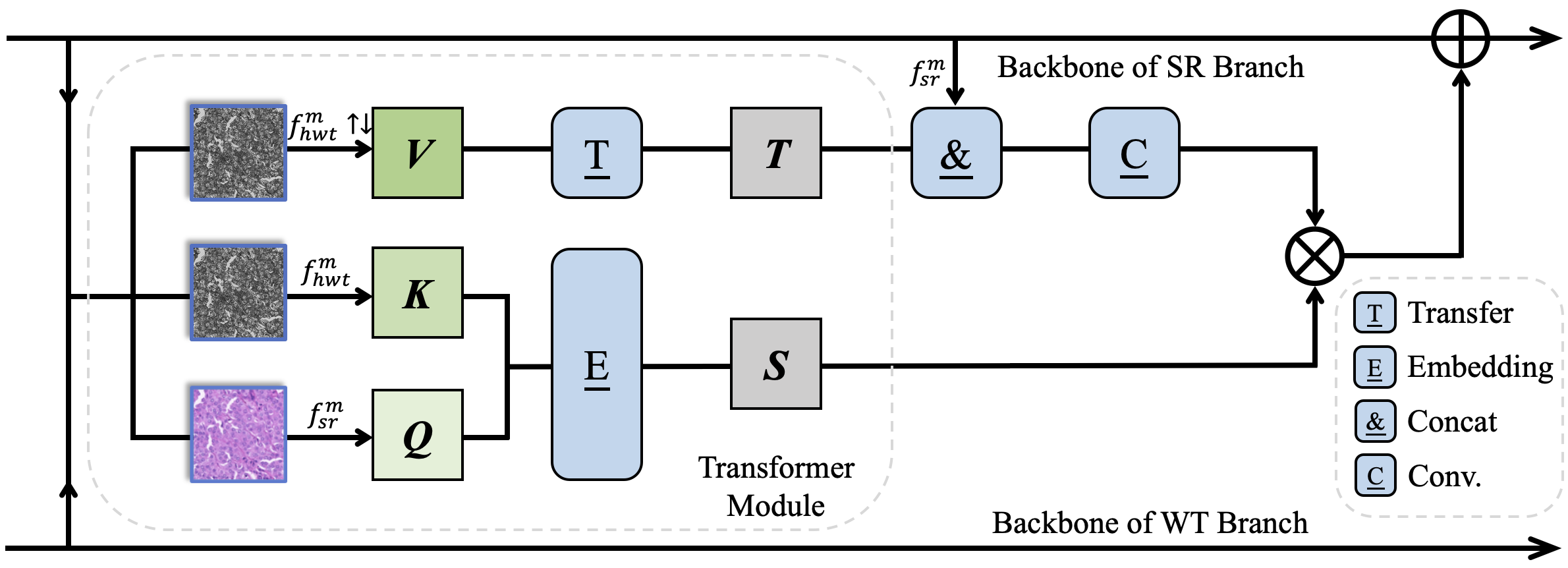}
\caption{The architecture of the Transformer block. In different network phases, features from two branches denoted Q, K, and V, enter the Transformer block. After transfer and embedding, these features are fused into the SR branch.}\label{fig3}
\end{figure*}

The WT branch is dedicated to capturing wavelet information and exhibits a superior capacity to generalize high-frequency details compared to the SR branch. Drawing inspiration from \cite{bib15}, we introduce the Transformer block to facilitate the SR branch in learning high-frequency information representation from the WT branch. In our Transformer block, the feature $f_{sr}^{m}$ from the SR branch and the feature $f_{hwt}^{m}$ from the WT branch serve as Query (Q) and Key (K), respectively. Additionally, we denote $f_{hwt}^{m}\uparrow\downarrow$ as the feature $f_{hwt}^{m}$ after upsampling and downsampling, representing it as Value (V). The Transformer block initially partitions these features into patches, and for any pair of patches in Q and K, it computes their similarity $r_{(i,j)}$ through inner product, estimating the similarity between $I_{LR}$ and $I_{GT}^{'}$. A higher inner product signifies a stronger correlation between the patches and a richer transferable high-frequency information. The expression for $r_{(i,j)}$ is defined as Eq. (8):

\begin{equation}
r_{i, j}=\left\langle\frac{q_{i}}{|| q_{i} \|}, \frac{k_{j}}{|| k_{j}||}\right\rangle,(\mathrm{i}, \mathrm{j}) \in\left[1, \frac{\mathrm{H}}{\mathrm{S}} \times \frac{\mathrm{W}}{\mathrm{S}}\right]
\end{equation}
where $q_{i}$ and $k_{j}$ denote any patches in Q and K, while H, W, and S have the same meaning as Eq. (6).

Subsequently, the diagonal structure features are transferred from the WT branch to the SR branch. To identify the feature at the most relevant position in K for any Q and minimize less relevant feature transfer, we obtain the index h and the attention graph S based on $r_{(i,j)}$. Specifically, h helps locate the most relevant high-frequency feature T in V, and $h_{i}$ represents the index of the most relevant patches for the ith patch in both $I_{GT}^{'}$ and $I_{LR}$. Meanwhile, S records the correlation between any patches from Q and their corresponding most relevant patches in K. We then fuse features from both branches, combining the joint Q and T and sending them through the convolutional layer C to generate the feature $f_{QT}$. This feature, $f_{QT}$, is further combined with the attention graph S to produce the feature $f_{QTS}$. By incorporating S, we enable the precise utilization of migrated high-frequency texture T, giving greater weight to highly relevant information and suppressing less relevant details. Finally, $f_{QTS}$ guides the SR branch task by uniting with the original SR branch information Q. The working process of the Transformer is described in Eq. (9):

\begin{equation}
f_{s r}^{m}=Q \bigoplus \mathrm{C}\{\operatorname{Concat}(Q, T)\} \bigotimes S
\end{equation} 
where $Concat(.)$ signifies the vector concatenation operation, $\bigotimes$ denotes element-wise multiplication, and m, $f_{sr}^{m}$, and $\bigoplus$ maintain the same meanings as in Eq. (2).

\subsection{Loss Function}\label{subsec3.5}

To ensure the retention of valuable wavelet information from cross-scale wavelet transform images, CWT-Net optimizes the network by minimizing a linear combination of losses from both branches, expressed in Eq. (10):

\begin{equation}
L_{C W T}=\lambda_{1} L_{S R}+\lambda_{2} L_{W T}
\end{equation} 
where $L_{CWT}$ represents the overall loss of CWT-Net, while $L_{SR}$ and $L_{WT}$ represent the losses of the SR branch and WT branch, respectively. $\lambda_{1}$ and $\lambda_{2}$ are the respective loss weights of the two branches. Given a training set {$I_{GT}$, $I_{GT}^{'}$, $I_{LR}$} with N samples and prediction results {$I_{HR}$, $I_{WT}$}, $L_{SR}$ and $L_{WT}$ can be further defined as Eq. (11) and Eq. (12):

\begin{equation}
L_{S R}=L_{L_{1}}\left(I_{H R}, I_{G T}\right)+\lambda_{3} L_{S S I M}\left(I_{H R}, I_{G T}\right) 
\end{equation} 
\begin{equation}
\begin{split}
L_{W T}=L_{L_{1}}\left(I_{WT}, D W T\left(I_{GT}^{\prime}\right)\right)
+\lambda_{4} L_{SSIM}\left(I_{WT}, D W T\left(I_{GT}^{\prime}\right)\right)
\end{split}
\end{equation}
where $L_{L_{1}}(.,.)$ signifies the $L_{1}$ loss, $L_{SSIM}(.,.)$ represents the SSIM loss \cite{bib26}, $I_{WT}$ denotes the output of the backbone network in the WT branch, $DWT(.)$ signifies the high-frequency features obtained after the Discrete Wavelet Transform (DWT) module, and $\lambda_{3}$ and $\lambda_{4}$ denote the weights of the SSIM loss in the respective branch losses.

In our optimization strategy, we primarily employ a combination of both L1 loss and SSIM loss rather than relying solely on one of them. The use of MSE-based optimization in some SR networks \cite{bib4,bib5,bib6}, especially with large upsampling multipliers, often results in a notable loss of high-frequency details in $I_{HR}$ and excessively smooth textures, despite yielding better performance in terms of PSNR and SSIM. Studies \cite{bib6} have shown that PSNR/SSIM metrics do not consistently correlate with subjective human visual evaluation. $L_{1}$ loss maintains brightness and color consistency and can tolerate larger errors compared to $L_{2}$ loss \cite{bib37}. Regarding texture preservation, $L_{1}$ loss outperforms MSE loss and $L_{2}$ loss, reproducing more high-frequency details. The $L_{1}$ loss is described as Eq. (13):

\begin{equation}
L_{L_{1}}=\frac{\sum_{i=1}^{r H} \sum_{j=1}^{r W}\left\|\theta_{G}\left(I_{L R}\right)^{(i, j)}-I_{GT}^{(i, j)}\right\|_{1}}{N^{2} \times H \times W}
\end{equation}
where N, H, and W represent the number of samples, pixel height, and pixel width of the samples, respectively, and $\theta_{G}(.)$ denotes the computational process of the network.

Structural Similarity (SSIM) measures image similarity based on luminance, contrast, and structure. Human vision is less sensitive to the absolute luminance/color of pixels, and SSIM loss better quantifies human visual perception by considering the location of edges and textures. The SSIM loss is defined as Eq. (14):

\begin{equation}
L_{S S I M}=1-\operatorname{SSIM}\left(\theta_{G}\left(I_{L R}\right), I_{GT}\right)
\end{equation}
where $SSIM(.,.)$ represents the structural similarity measure between the two samples.

In our experiments, we set $(\lambda_{1},\lambda_{2})=(0.3,0.7)$. We assigned a higher weight to the loss in the WT branch compared to the SR branch, and this choice was driven by several considerations. First, the SR branch benefits from complete input features, while the WT branch only processes high-frequency features from the wavelet transform. Consequently, the SR branch exhibits higher robustness. Second, in SR tasks with a large upsampling factor, the WT branch dynamically increases the depth of the backbone network, potentially introducing more errors. Lastly, although the SR branch addresses our primary task, the optimization outcomes of the WT branch significantly impact the SR branch's results. We set $(\lambda_{3},\lambda_{4})=(0.2,0.2)$. The combination of $L_{L_{1}}$ as the primary component of the loss function ensures better convergence and maintains visual quality for the SR task. On the other hand, $L_{SSIM}$ enhances pixel-level accuracy, balances the dynamic range of predictions, and stabilizes the model optimization by amplifying differences between prediction results and training samples.

\section{Experiment}\label{sec4}

\subsection{Dataset}\label{subsec4.1}

WSI refers to the process of scanning an entire microscope slide and creating a single high-resolution digital file, primarily applied in the field of pathology cell images. Typically, the pixel sizes between two adjacent sampling levels in WSI are spaced at twice the distance to enable rapid and accurate downsampling.

The CAMELYON dataset \cite{bib38} comprises 400 WSI images of breast cancer anterior lymph node sections. Li Y {\it et al.} \cite{bib33} provided pre-sampled coordinate pairs and five levels of sampling resolution for cancer and non-cancer regions within the CAMELYON dataset. To introduce realistic cross-scale information to CWT-Net, we selected a total of 1200 coordinates from both regions and sampled them at $40\times$ magnification (0.243 microns per pixel), $20\times$ magnification (0.486 microns per pixel), and $10\times$ magnification (0.972 microns per pixel) across patch sizes of $1024\times1024$, $512\times512$, and $256\times256$, respectively. These images were randomly organized and split into a training set and a test set at a 5:1 ratio. We named this dataset MLCamSR to facilitate the super-resolution task for multi-level histopathology images. While constructing MLCamSR, we carefully assessed the quality of WSI images, removed samples with significant blank areas, and ensured that the RGB mean values of all samples fell within the appropriate range (RGB mean values were set to [0.7204,0.4298,0.6379] for the training set).

Sun K {\it et al.} \cite{bib39} contributed a WSI image dataset encompassing 10 human body systems. Pathology slides were obtained from Xiangya Hospital of Central South University. Technicians randomly selected original WSI files representing typical diseases from each human system, which were then medically reviewed by two pathologists. Slides were scanned using a digital pathology scanner to obtain samples at $40\times$ magnification. Samples at $20\times$ magnification, $10\times$ magnification, and $5\times$ magnification were derived from $40\times$ magnification images by applying the bicubic algorithm three times consecutively. The patch sizes ranged from $1024\times1024$ to $128\times128$. We conducted experiments using a subset of \cite{bib39}, which consists of 1200 samples and follows the same structure as MLCamSR, and we named it FTMS.

The PCam dataset \cite{bib40}, a subset of \cite{bib38}, encompasses 327,680 patches sampled from $10\times$ magnification images, each measuring $96\times96$.

\subsection{Experimental details}\label{subsec4.2}

We utilize CWT-Net as our baseline model, consisting of 12 CWTBs. In this model, all convolutional layers employ a $3\times3$ filter size with 64 channels, and the WR module reduces channels by a factor of 8. During training, we partition training samples into $64\times64$ patches for the SR branch and apply random 90-degree rotations to augment the data. Each $B_{RB_{w}}$ consists of 2 RBs and the WR module contains 4 WRBs, when the upsampling factor is 2. For each subsequent increase in the upsampling factor by a factor of 1, 1 RB is added to each $B_{RB_{w}}$,  and 2 WRBs are added to the WR module.

Our network is implemented on the torch 1.12.1 platform, with one NVIDIA GeForce RTX 3090 (24GB) and one NVIDIA TITAN Xp (12GB) for all experiments. In the quantitative evaluation (Section \ref{subsec4.3}), for the $2\times$ upsampling task, we use the original HR image as input to the WT branch. For other tasks, the WT branch input is obtained from images one sampling level higher than the SR branch input. If the WR module is used solely during testing, CWT-Net initializes WR module weights from a standard normal distribution. All ablation experiments (Section \ref{subsec4.5}) are based on the MLCamSR $2\times$ upsampling task, and CWT-Net is trained for 1000 batches (100 epochs) in all experiments, unless specified otherwise. In the classification task evaluation (Section \ref{subsec4.6}), we segment each MLCamSR image into $128\times128$ images to match the classification network's feature size. These segmented images retain the same category labels as the original images.

\subsection{Quantitative Experiments}\label{subsec4.3}

\begin{table*}[ht]
\footnotesize  
\begin{center}
\caption{Quantitative results on the two datasets at various scaling factors. The best results are underlined.}\label{tab1}
\setlength{\tabcolsep}{4.5mm}
\begin{tabular}{cccccccc}
\toprule%
Dataset & Method & \multicolumn{2}{c}{2$\times$} & \multicolumn{2}{c}{4$\times$} & \multicolumn{2}{c}{8$\times$} \\
\cmidrule{3-4}\cmidrule{5-6}\cmidrule{7-8}
 & & PSNR(dB) & SSIM & PSNR(dB) & SSIM & PSNR(dB) & SSIM \\
\midrule
MLCamSR & bicubic & 35.247 & .9427 & 27.019 & .6659 & 22.475 & .2276 \\
 & SRCNN\cite{bib4} & 35.522 & .9427 & 27.475 & .7329 & 22.489 & .3624 \\ 
 & SRGAN\cite{bib6} & 35.649 & .9518 & 27.626 & .7406 & 22.946 & .3590 \\
 & EDSR\cite{bib9} & 38.190 & .9677 & 27.886 & .7479 & 23.007 & .4912 \\
 & RCAN\cite{bib10} & 38.327 & .9692 & 28.339 & .7887 & 23.075 & .4894 \\
 & MWCNN\cite{bib13} & 38.800 & .9722 & 28.941 & .8130 & 23.590 & .5373 \\
 & SWD-Net\cite{bib27} & 39.228 & .9719 & 29.349 & .8239 & 23.983 & .5471 \\
 & CWT-Net & \underline{39.333} & \underline{.9797} & \underline{29.639} & \underline{.8304} & \underline{24.140} & \underline{.5535} \\
\midrule
FTMS & bicubic & 27.176 & .6613 & 23.694 & .4814 & 20.554 & .2696 \\
 & SRCNN\cite{bib4} & 27.477 & .6620 & 23.810 & .4802 & 20.592 & .2707 \\
 & SRGAN\cite{bib6} & 27.939 & .6621 & 23.847 & .4849 & 20.001 & .2681 \\
 & EDSR\cite{bib9} & 28.145 & .6650 & 24.557 & .4727 & 20.375 & .2730 \\
 & RCAN\cite{bib10} & 28.255 & .6698 & 24.639 & .4750 & 20.029 & .2722 \\
 & MWCNN\cite{bib13} & 28.458 & .7443 & 24.992 & .4907 & 20.767 & \underline{.2762} \\
 & SWD-Net\cite{bib27} & 28.586 & .7602 & 25.171 & .5014 & 20.874 & .2750 \\ 
 & CWT-Net & \underline{28.972} & \underline{.7783} & \underline{25.237} & \underline{.5050} & \underline{20.885} & .2701 \\
\bottomrule
\end{tabular}
\end{center}
\end{table*}

Table.\ref{tab1} presents objective metric results for various SR methods across $2\times$, $4\times$, and $8\times$ tasks. CWT-Net excels in performance on both datasets. In the $2\times$ task, CWT-Net surpasses competitors by 0.105 dB/0.0078 and 0.386 dB/0.0181 in objective metrics. This success is attributed to the varying cross-scale information in FTMS images containing multiple human tissue cells, favoring the WT branch and WR modules.

For the $4\times$ task, CWT-Net outperforms the second-place method by 0.290 dB/0.0065 and 0.066 dB/0.0036 in terms of objective metrics, respectively. In SR tasks, as magnification increases, the contribution of $I_{GT}^{'}$ and $I_{LR}$ diminishes, making CWT-Net more reliant on cross-scale information quality. This trend continues in the 8x task, where CWT-Net gains 0.157 dB/0.0064 in objective metrics on MLCamSR but shows limited improvement on FTMS, with no substantial enhancement in structural similarity.

In summary, CWT-Net consistently outperforms FTMS using MLCamSR in all tasks. MLCamSR's utilization of actual sampling processes for cross-scale information, as opposed to image generation algorithms, underscores CWT-Net's ability to leverage and discern differences between cross-scale data effectively.

\subsection{Qualitative Experiments}\label{subsec4.4}

\begin{figure*}[htbp]%
\centering
\includegraphics[width=0.8\textwidth]{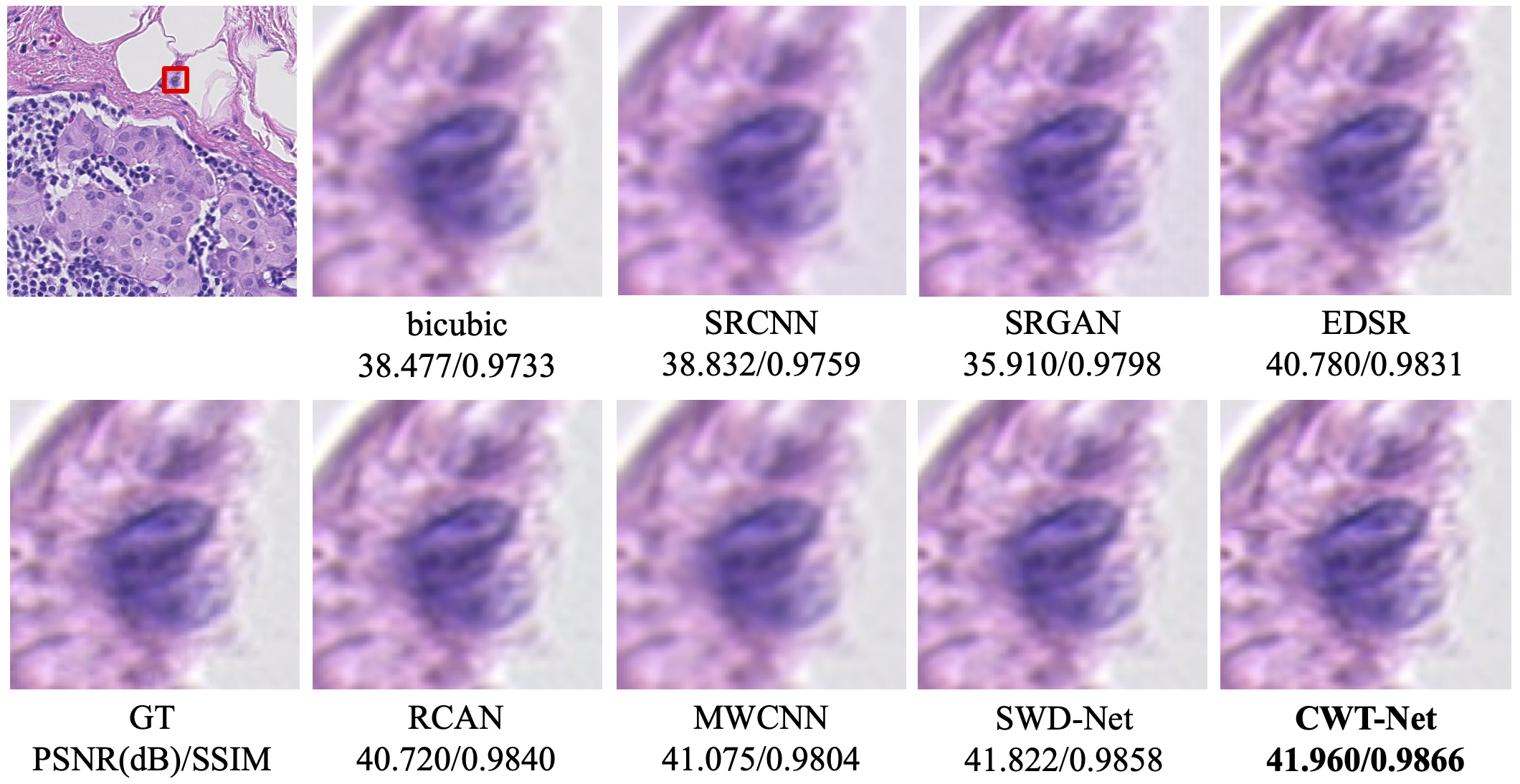}
\caption{Qualitative results in the $2\times$ up-sampling task.}\label{fig4}
\end{figure*}

\begin{figure*}[htbp]%
\centering
\includegraphics[width=0.8\textwidth]{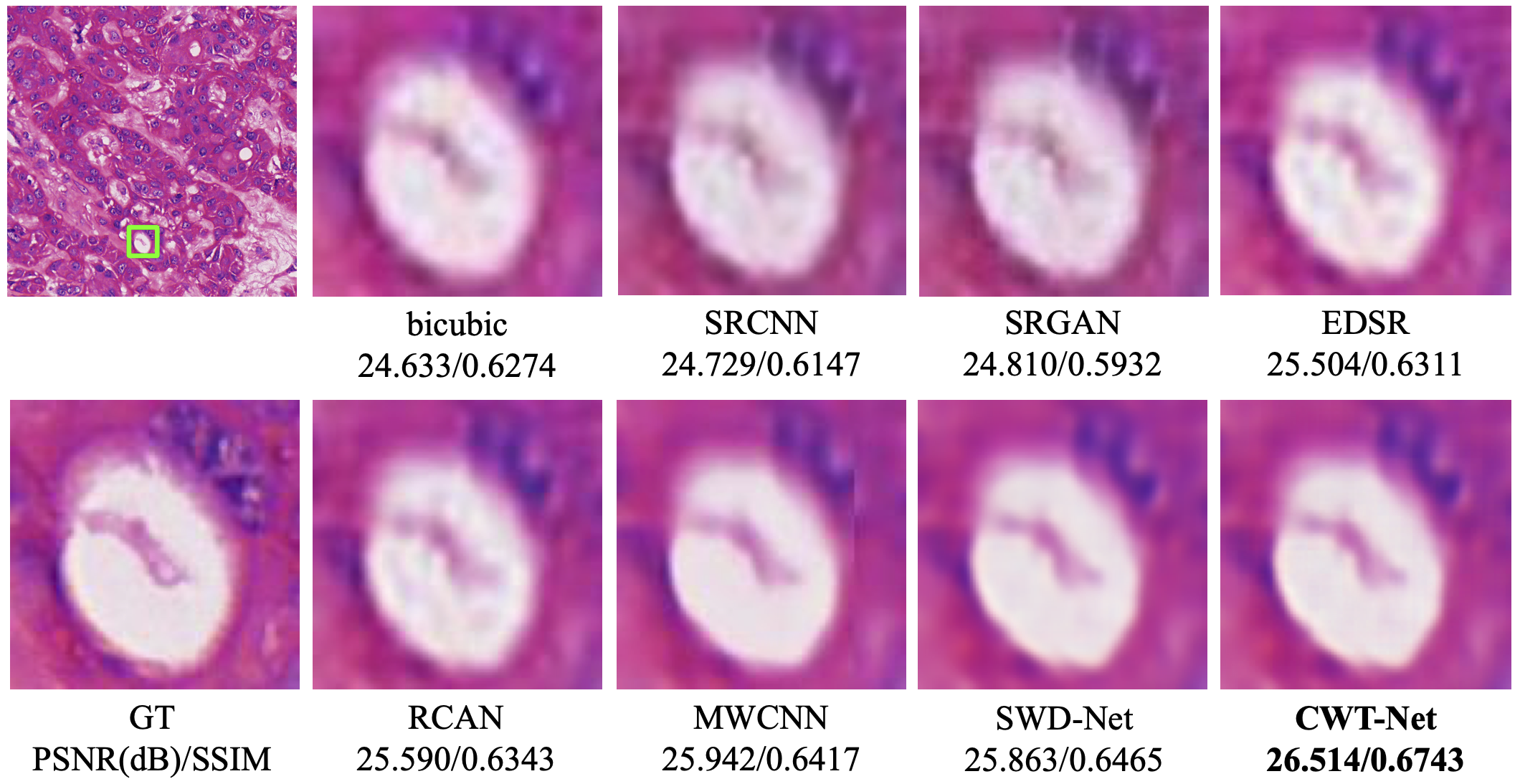}
\caption{Qualitative results in the $4\times$ up-sampling task.}\label{fig5}
\end{figure*}

Fig.\ref{fig4} and Fig.\ref{fig5} offer visual representations of $2\times$ and $4\times$ tasks, with CWT-Net notably exhibiting greater image detail compared to other models. This difference is particularly pronounced in the high-frequency structural information present in the input images.

\subsection{Ablation Experiments}\label{subsec4.5}

\subsubsection{Effect of network components on model performance}\label{subsubsec4.5.1}

\begin{table}[ht]
\footnotesize
\begin{center}
\caption{The influence of components in CWT-Net on the benchmark model's outcomes.}\label{tab2}
\setlength{\tabcolsep}{1.0mm}
\begin{tabular}{ccccc}
\toprule%
SR B & WT B/Transformer & DWT & WR & PSNR(dB)/SSIM\\
\midrule
\checkmark & & & & 39.161/.9712 \\
\checkmark & \checkmark & & & 39.145/.9619 \\
\checkmark & \checkmark & \checkmark & & 38.928/.9683 \\
\checkmark & \checkmark & \checkmark & \checkmark & 39.333/.9797 \\
\bottomrule
\end{tabular}
\end{center}
\end{table}

CWT-Net is divided into two branches, with the WT branch further categorized into three components based on the flow of features: the backbone network and Transformer module, the DWT module, and the WR module. The results of the ablation experiments are presented in Table.\ref{tab2}.

The SR branch is capable of independently conducting feature extraction and amplification. However, its performance closely approximates that of a single primary network, similar in construction to EDSR. Upon the introduction of the backbone network and Transformer module into the WT branch, there is a slight decrease in PSNR, from 39.161 to 39.145. Clearly, the use of the Transformer module alone does not provide beneficial features for the SR branch.

However, with the inclusion of the DWT module, PSNR decreases once more. Nevertheless, the SSIM metric increases from 0.9619 to 0.9683, demonstrating that wavelet transform features can provide structural information essential for the SR task. Finally, with the addition of the WR module, PSNR/SSIM increases to 39.333/0.9797, affirming the superior performance of the WR module.

\subsubsection{Effect of input feature of WT branches on model performance}\label{subsubsec4.5.2}

\begin{table}[ht]
\footnotesize
\begin{center}
\caption{Investigation of input features of the WT branch.}
\label{tab3}
\begin{tabular}{ccc}
\toprule
Features & LPIPS\cite{bib41}: AlexNet/VGG & PSNR(dB)/SSIM \\
\midrule
$I_{GT}$ & .00832/.01679 & 44.035/.9933 \\
$I_{GT}^{'}$(ours) & .04091/.05131 & 39.333/.9797 \\
$I_{GT}\uparrow\downarrow$ & .04275/.05209 & 39.042/.9749 \\
$I_{LR}$ & .06327/.07900 & 38.670/.9680 \\
\bottomrule
\end{tabular}
\end{center}
\end{table}

The WT branches are highly sensitive to the degradation of their input features. We utilize different features as initial inputs for the backbone network in the WT branch, as illustrated in Table.\ref{tab3}. To further assess the subjective perceptual quality of these strategies, we employ the Learned Perceptual Image Patch Similarity (LPIPS) metric \cite{bib41}. LPIPS extracts image features and computes similarities by pre-training neural networks (such as AlexNet, VGG, or others) to mimic human subjective evaluations. The closer the LPIPS score is to 0, the more similar the two input images are perceived by humans.

In our data partitioning strategy, $I_{GT}$ contains the most complete high-frequency information, offering the model the highest potential gain from the WT branch. The strategy of using $I_{GT}^{'}$ aligns with the benchmark model. When $I_{GT}$ undergoes sequential upsampling and downsampling, denoted as $I_{GT}\uparrow\downarrow$, the loss of high-frequency information becomes significantly greater than the loss of cross-scale images. Opting for $I_{LR}$ allows CWT-Net to conform closely to the SISR paradigm, but it introduces overlapping artifacts into the final results. As observed, our chosen strategy brings the model closest to achieving performance akin to the theoretical limit across all metrics.

\subsubsection{Effect of the loss function and its optimization strategy} \label{subsubsec4.5.3}

\begin{figure}[htbp]%
\centering
\includegraphics[width=0.48\textwidth]{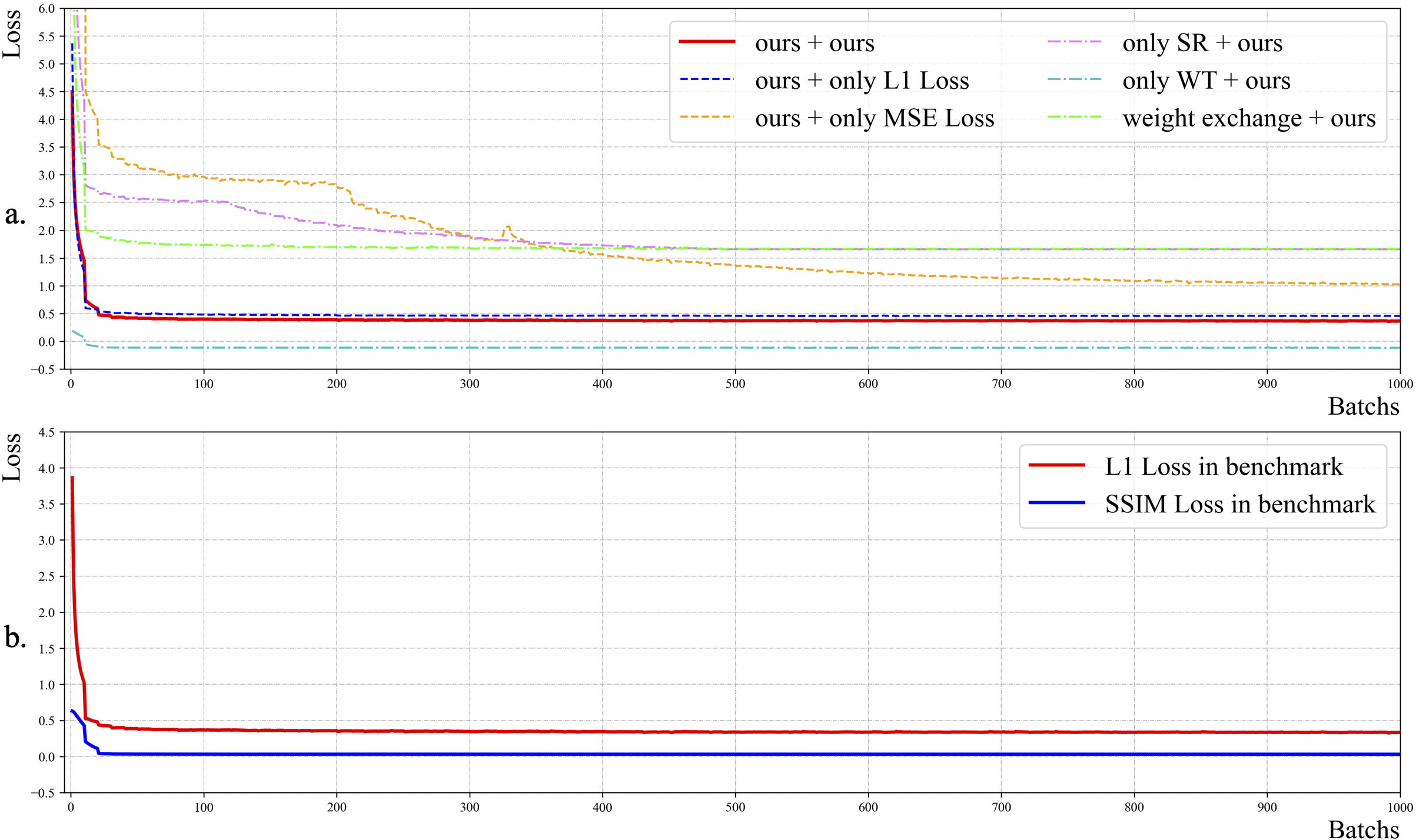}
\caption{Visualization of network optimization. a. The performance of different combinations of loss functions and optimization strategies. The red line (ours+ours) represents our benchmark model. b. Performance of $L_{1}$ loss and SSIM loss in the benchmark model.}\label{fig6}
\end{figure}

\begin{table}[ht]
\footnotesize
\begin{center}
\caption{Investigation of the loss function and its optimization strategy.}
\setlength{\tabcolsep}{1.5mm}
\label{tab4}
\begin{tabular}{ccc}
\toprule%
Loss Function & Optimization Strategy & PSNR(dB)/SSIM \\
\midrule
ours & ours & 39.333/.9797 \\
only $L_{1}$ Loss & ours & 39.077/.9714 \\
only MSE Loss & ours & 39.030/.9647 \\
ours & only SR & 38.895/.9699 \\
ours & only WT & 9.947/.0658 \\
ours & weight exchange & 39.286/.9797 \\
\bottomrule
\end{tabular}
\end{center}
\end{table}

Table.\ref{tab4} displays the objective metric results of CWT-Net under various loss functions and optimization strategies. It's evident that relying solely on either the $L_{1}$ loss or MSE loss can negatively impact performance. "only SR": The loss function solely computes the dissimilarity between the output of the SR branch and the original HR image. Clearly, optimizing only the SR branch enables CWT-Net to perform the SR task to some extent, but with significantly lower performance than optimizing both branches simultaneously. "only WT": The loss function solely computes the dissimilarity between the output of the WT branch and the original wavelet features. Optimizing only the WT branch does not lead to model convergence. "weight exchange": The loss weights $\lambda_{1}$ and $\lambda_{2}$ of the two branches in the baseline model are swapped. When the weights of the two branches are exchanged, the model achieves performance closest to that of the baseline model, but still with some disparity. Evidently, our strategy allows CWT-Net to attain peak performance.

In Fig.\ref{fig6}.a, we provide a visualization of the optimization combinations during the training process, corresponding to the data presented in Table.\ref{tab4}. When comparing it to the loss function strategy we propose, the curve using only $L_{1}$ loss shows more pronounced oscillations in the early stages of the training process. The convergence rate of the MSE loss is slower compared to that of the $L_{1}$ loss. The SSIM loss introduces a smoothing effect in CWT-Net by capturing local features through a sliding window, which aligns with the subjective perception of human vision. Our proposed loss function strategy maintains high stability throughout the training phase (see Fig.\ref{fig5}.b), while relying solely on $L_{1}$ loss results in the loss of some structural information. It's important to note that the network optimization strategy has a more significant impact on the baseline model compared to the loss function strategy.

\subsubsection{Effect of the number of CWTBs on model performance} \label{subsubsec4.5.4} 

\begin{table}[ht]
\footnotesize
\begin{center}
\caption{Investigation of the number of CWTB.}
\setlength{\tabcolsep}{5.0mm}
\label{tab5}
\begin{tabular}{cccc}
\toprule
Number & PSNR(dB)/SSIM & Avg. Time\\
\midrule
2 & 39.054/.9714 & 6.7 \\
4 & 39.067/.9711 & 11.1 \\
8 & 39.219/.9723 & 20.5 \\
12 & 39.278/.9748 & 30.3 \\
24 & 39.577/.9799 & 56.5 \\
\bottomrule
\end{tabular}
\end{center}
\end{table}

Our benchmark model incorporates 12 CWTBs. Table.\ref{tab5} illustrates the performance of CWT-Net with varying network depths. In a training scenario comprising 500 Batches, we observe that network performance improves with increased depth, underscoring the effectiveness of our underlying substructure, the CWTB. 

However, it's important to note that greater network depth also leads to increased time costs. For instance, when there are 8 CWTBs, the network achieves a PSNR of 99.85\% compared to the benchmark model, with a time cost of 54.15\%. Conversely, with 24 CWTBs, the network achieves a PSNR of 100.76\% compared to the benchmark model, but the time cost increases significantly to 186.47\%. Therefore, in practical medical diagnosis scenarios, medical technicians may need to consider the computational cost of histopathology images and the desired accuracy threshold for diagnosis. In some cases, using a lightweight model could be a more efficient choice to streamline the diagnostic process.

\subsection{Evaluation of image classification tasks for diagnosis}\label{subsec4.6}

\begin{table}[ht]
\footnotesize
\begin{center}
\caption{Results on the binary classification task of breast cancer pathology images using ResNet-50.}
\setlength{\tabcolsep}{6.0mm}
\label{tab6}
\begin{tabular}{ccc}
\toprule%
Method & \multicolumn{2}{c}{Acc}\\
\cmidrule{2-3}
 & MLCamSR & PCam \\
\midrule
GT & 91.83 & 90.03 \\
bicubic(↓) & 62.89 & 63.22 \\
bicubic & 79.53 & 66.40 \\
SRCNN\cite{bib4} & 77.76 & 68.87 \\
SRGAN\cite{bib6} & 81.04 & 70.35 \\
EDSR\cite{bib9} & 82.72 & 74.19 \\
RCAN\cite{bib10} & 84.31 & 79.68 \\
MWCNN\cite{bib13} & 82.71 & 76.42 \\
SWD-Net\cite{bib27} & 86.47 & 80.48 \\
CWT-Net & 88.12 & 80.85 \\
\bottomrule
\end{tabular}
\end{center}
\end{table} 

To further assess the impact of CWT-Net on diagnostic tasks, we employed ResNet-50 \cite{bib8} as the baseline classifier for histopathology image classification using the results from the $2\times$ SR task on the MLCamSR and PCam datasets. The experiment results are presented in Table.\ref{tab6}. Both deep learning-based SR models outperformed bicubic interpolation. CWT-Net achieved the highest classification performance, with improvements of 7.59\% and 14.45\% in accuracy over bicubic interpolation and 1.65\% and 0.37\% over the second-place SWDNet. These results highlight the significant contribution of CWT-Net to enhancing histopathology image classification accuracy and its potential utility in medical diagnosis.

\section{Conclusion}\label{sec5}

In this research, we introduce CWT-Net, a SR model that leverages cross-scale wavelet features and a Transformer structure. CWT-Net is applied to the task of upsampling histopathology images from various human systems, with a primary focus on enhancing the reconstruction of structural information. We explore the advantages of the RefSR paradigm. CWT-Net comprises three core components: the SR branch for upsampling, the WT branch for generating wavelet transform features from cross-scale images, and the Transformer module for transmitting structural information to the SR branch. The WR module is specially designed to support the RefSR paradigm within the SISR task. To enable training with undegraded cross-scale information, we create the MLCamSR benchmark dataset. CWT-Net exhibits state-of-the-art performance across various datasets and upsampling scales, enhancing image classification networks in diagnostics. A series of ablation studies confirm the robustness and effectiveness of CWT-Net. In the future, we anticipate using CWT-Net and its substructures to provide compatible pre-training measures or priors for other histopathology image-related tasks, opening new research and application possibilities in the field.


\bibliographystyle{IEEEtran}
\bibliography{ref}

\begin{thebibliography}{10}
\providecommand{\url}[1]{#1}
\csname url@samestyle\endcsname
\providecommand{\newblock}{\relax}
\providecommand{\bibinfo}[2]{#2}
\providecommand{\BIBentrySTDinterwordspacing}{\spaceskip=0pt\relax}
\providecommand{\BIBentryALTinterwordstretchfactor}{4}
\providecommand{\BIBentryALTinterwordspacing}{\spaceskip=\fontdimen2\font plus
\BIBentryALTinterwordstretchfactor\fontdimen3\font minus \fontdimen4\font\relax}
\providecommand{\BIBforeignlanguage}[2]{{%
\expandafter\ifx\csname l@#1\endcsname\relax
\typeout{** WARNING: IEEEtran.bst: No hyphenation pattern has been}%
\typeout{** loaded for the language `#1'. Using the pattern for}%
\typeout{** the default language instead.}%
\else
\language=\csname l@#1\endcsname
\fi
#2}}
\providecommand{\BIBdecl}{\relax}
\BIBdecl

\bibitem{bib1}
Z.~Wang, J.~Chen, and S.~C. Hoi, ``Deep learning for image super-resolution: A survey,'' \emph{IEEE transactions on pattern analysis and machine intelligence}, vol.~43, no.~10, pp. 3365--3387, 2020.

\bibitem{bib2}
M.~D. Robinson, C.~A. Toth, J.~Y. Lo, and S.~Farsiu, ``Efficient fourier-wavelet super-resolution,'' \emph{IEEE Transactions on Image Processing}, vol.~19, no.~10, pp. 2669--2681, 2010.

\bibitem{bib3}
H.~Ji and C.~Ferm{\"u}ller, ``Robust wavelet-based super-resolution reconstruction: theory and algorithm,'' \emph{IEEE Transactions on Pattern Analysis and Machine Intelligence}, vol.~31, no.~4, pp. 649--660, 2008.

\bibitem{bib4}
C.~Dong, C.~C. Loy, K.~He, and X.~Tang, ``Image super-resolution using deep convolutional networks,'' \emph{IEEE transactions on pattern analysis and machine intelligence}, vol.~38, no.~2, pp. 295--307, 2015.

\bibitem{bib5}
C.~Dong, C.~C. Loy, and X.~Tang, ``Accelerating the super-resolution convolutional neural network,'' in \emph{Computer Vision--ECCV 2016: 14th European Conference, Amsterdam, The Netherlands, October 11-14, 2016, Proceedings, Part II 14}.\hskip 1em plus 0.5em minus 0.4em\relax Springer, 2016, pp. 391--407.

\bibitem{bib6}
C.~Ledig, L.~Theis, F.~Husz{\'a}r, J.~Caballero, A.~Cunningham, A.~Acosta, A.~Aitken, A.~Tejani, J.~Totz, Z.~Wang \emph{et~al.}, ``Photo-realistic single image super-resolution using a generative adversarial network,'' in \emph{Proceedings of the IEEE conference on computer vision and pattern recognition}, 2017, pp. 4681--4690.

\bibitem{bib8}
K.~He, X.~Zhang, S.~Ren, and J.~Sun, ``Deep residual learning for image recognition,'' in \emph{Proceedings of the IEEE conference on computer vision and pattern recognition}, 2016, pp. 770--778.

\bibitem{bib9}
B.~Lim, S.~Son, H.~Kim, S.~Nah, and K.~Mu~Lee, ``Enhanced deep residual networks for single image super-resolution,'' in \emph{Proceedings of the IEEE conference on computer vision and pattern recognition workshops}, 2017, pp. 136--144.

\bibitem{bib10}
Y.~Zhang, K.~Li, K.~Li, L.~Wang, B.~Zhong, and Y.~Fu, ``Image super-resolution using very deep residual channel attention networks,'' in \emph{Proceedings of the European conference on computer vision (ECCV)}, 2018, pp. 286--301.

\bibitem{bib11}
Y.~Zhang, Y.~Tian, Y.~Kong, B.~Zhong, and Y.~Fu, ``Residual dense network for image super-resolution,'' in \emph{Proceedings of the IEEE conference on computer vision and pattern recognition}, 2018, pp. 2472--2481.

\bibitem{bib12}
Y.~Lu, Y.-W. Tai, and C.-K. Tang, ``Attribute-guided face generation using conditional cyclegan,'' in \emph{Proceedings of the European conference on computer vision (ECCV)}, 2018, pp. 282--297.

\bibitem{bib13}
P.~Liu, H.~Zhang, K.~Zhang, L.~Lin, and W.~Zuo, ``Multi-level wavelet-cnn for image restoration,'' in \emph{Proceedings of the IEEE conference on computer vision and pattern recognition workshops}, 2018, pp. 773--782.

\bibitem{bib31}
H.~Huang, R.~He, Z.~Sun, and T.~Tan, ``Wavelet-srnet: A wavelet-based cnn for multi-scale face super resolution,'' in \emph{Proceedings of the IEEE international conference on computer vision}, 2017, pp. 1689--1697.

\bibitem{bib44}
J.~Lei, Z.~Zhang, X.~Fan, B.~Yang, X.~Li, Y.~Chen, and Q.~Huang, ``Deep stereoscopic image super-resolution via interaction module,'' \emph{IEEE Transactions on Circuits and Systems for Video Technology}, vol.~31, no.~8, pp. 3051--3061, 2020.

\bibitem{bib45}
H.~Wang, S.~Li, and M.~Zhao, ``A lightweight recurrent aggregation network for satellite video super-resolution,'' \emph{IEEE Journal of Selected Topics in Applied Earth Observations and Remote Sensing}, vol.~17, pp. 685--695, 2024.

\bibitem{bib47}
H.~Sheng, S.~Wang, D.~Yang, R.~Cong, Z.~Cui, and R.~Chen, ``Cross-view recurrence-based self-supervised super-resolution of light field,'' \emph{IEEE Transactions on Circuits and Systems for Video Technology}, pp. 1--1, 2023.

\bibitem{bib48}
Z.~Wu, W.~Liu, J.~Li, C.~Xu, and D.~Huang, ``Sfhn: Spatial-frequency domain hybrid network for image super-resolution,'' \emph{IEEE Transactions on Circuits and Systems for Video Technology}, vol.~33, no.~11, pp. 6459--6473, 2023.

\bibitem{bib49}
Y.~Zuo, J.~Xie, H.~Wang, Y.~Fang, D.~Liu, and W.~Wen, ``Gradient-guided single image super-resolution based on joint trilateral feature filtering,'' \emph{IEEE Transactions on Circuits and Systems for Video Technology}, vol.~33, no.~2, pp. 505--520, 2023.

\bibitem{bib50}
D.~Cheng, L.~Chen, C.~Lv, L.~Guo, and Q.~Kou, ``Light-guided and cross-fusion u-net for anti-illumination image super-resolution,'' \emph{IEEE Transactions on Circuits and Systems for Video Technology}, vol.~32, no.~12, pp. 8436--8449, 2022.

\bibitem{bib51}
Z.~Liu, Z.~Li, X.~Wu, Z.~Liu, and W.~Chen, ``Dsrgan: Detail prior-assisted perceptual single image super-resolution via generative adversarial networks,'' \emph{IEEE Transactions on Circuits and Systems for Video Technology}, vol.~32, no.~11, pp. 7418--7431, 2022.

\bibitem{bib52}
A.~Niu, Y.~Zhu, C.~Zhang, J.~Sun, P.~Wang, I.~S. Kweon, and Y.~Zhang, ``Ms2net: Multi-scale and multi-stage feature fusion for blurred image super-resolution,'' \emph{IEEE Transactions on Circuits and Systems for Video Technology}, vol.~32, no.~8, pp. 5137--5150, 2022.

\bibitem{bib53}
J.~Zhang, C.~Long, Y.~Wang, H.~Piao, H.~Mei, X.~Yang, and B.~Yin, ``A two-stage attentive network for single image super-resolution,'' \emph{IEEE Transactions on Circuits and Systems for Video Technology}, vol.~32, no.~3, pp. 1020--1033, 2022.

\bibitem{bib20}
X.~Zhao, Y.~Zhang, T.~Zhang, and X.~Zou, ``Channel splitting network for single mr image super-resolution,'' \emph{IEEE transactions on image processing}, vol.~28, no.~11, pp. 5649--5662, 2019.

\bibitem{bib21}
C.~Qiao, D.~Li, Y.~Guo, C.~Liu, T.~Jiang, Q.~Dai, and D.~Li, ``Evaluation and development of deep neural networks for image super-resolution in optical microscopy,'' \emph{Nature Methods}, vol.~18, no.~2, pp. 194--202, 2021.

\bibitem{bib22}
Z.~Li, Q.~Liu, Y.~Li, Q.~Ge, Y.~Shang, D.~Song, Z.~Wang, and J.~Shi, ``A two-stage multi-loss super-resolution network for arterial spin labeling magnetic resonance imaging,'' in \emph{Medical Image Computing and Computer Assisted Intervention--MICCAI 2019: 22nd International Conference, Shenzhen, China, October 13--17, 2019, Proceedings, Part III 22}.\hskip 1em plus 0.5em minus 0.4em\relax Springer, 2019, pp. 12--20.

\bibitem{bib23}
L.~Mukherjee, H.~D. Bui, A.~Keikhosravi, A.~Loeffler, and K.~W. Eliceiri, ``Super-resolution recurrent convolutional neural networks for learning with multi-resolution whole slide images,'' \emph{Journal of biomedical optics}, vol.~24, no.~12, pp. 126\,003--126\,003, 2019.

\bibitem{bib27}
Z.~Chen, X.~Guo, C.~Yang, B.~Ibragimov, and Y.~Yuan, ``Joint spatial-wavelet dual-stream network for super-resolution,'' in \emph{Medical Image Computing and Computer Assisted Intervention--MICCAI 2020: 23rd International Conference, Lima, Peru, October 4--8, 2020, Proceedings, Part V 23}.\hskip 1em plus 0.5em minus 0.4em\relax Springer, 2020, pp. 184--193.

\bibitem{bib42}
X.~Wu, Z.~Chen, C.~Peng, and X.~Ye, ``Mmsrnet: Pathological image super-resolution by multi-task and multi-scale learning,'' \emph{Biomedical Signal Processing and Control}, vol.~81, p. 104428, 2023.

\bibitem{bib54}
H.~Wang, X.~Hu, X.~Zhao, and Y.~Zhang, ``Wide weighted attention multi-scale network for accurate mr image super-resolution,'' \emph{IEEE Transactions on Circuits and Systems for Video Technology}, vol.~32, no.~3, pp. 962--975, 2022.

\bibitem{bib46}
Z.~Qiu, Y.~Hu, X.~Chen, D.~Zeng, Q.~Hu, and J.~Liu, ``Rethinking dual-stream super-resolution semantic learning in medical image segmentation,'' \emph{IEEE Transactions on Pattern Analysis and Machine Intelligence}, vol.~46, no.~1, pp. 451--464, 2024.

\bibitem{bib16}
Z.~Lu, J.~Li, H.~Liu, C.~Huang, L.~Zhang, and T.~Zeng, ``Transformer for single image super-resolution,'' in \emph{Proceedings of the IEEE/CVF conference on computer vision and pattern recognition}, 2022, pp. 457--466.

\bibitem{bib17}
J.~Liang, J.~Cao, G.~Sun, K.~Zhang, L.~Van~Gool, and R.~Timofte, ``Swinir: Image restoration using swin transformer,'' in \emph{Proceedings of the IEEE/CVF international conference on computer vision}, 2021, pp. 1833--1844.

\bibitem{bib24}
F.~Shahidi, ``Breast cancer histopathology image super-resolution using wide-attention gan with improved wasserstein gradient penalty and perceptual loss,'' \emph{IEEE Access}, vol.~9, pp. 32\,795--32\,809, 2021.

\bibitem{bib28}
C.-M. Feng, Y.~Yan, H.~Fu, L.~Chen, and Y.~Xu, ``Task transformer network for joint mri reconstruction and super-resolution,'' in \emph{Medical Image Computing and Computer Assisted Intervention--MICCAI 2021: 24th International Conference, Strasbourg, France, September 27--October 1, 2021, Proceedings, Part VI 24}.\hskip 1em plus 0.5em minus 0.4em\relax Springer, 2021, pp. 307--317.

\bibitem{bib29}
W.~Shi, J.~Caballero, F.~Husz{\'a}r, J.~Totz, A.~P. Aitken, R.~Bishop, D.~Rueckert, and Z.~Wang, ``Real-time single image and video super-resolution using an efficient sub-pixel convolutional neural network,'' in \emph{Proceedings of the IEEE conference on computer vision and pattern recognition}, 2016, pp. 1874--1883.

\bibitem{bib30}
J.~Kim, J.~K. Lee, and K.~M. Lee, ``Accurate image super-resolution using very deep convolutional networks,'' in \emph{Proceedings of the IEEE conference on computer vision and pattern recognition}, 2016, pp. 1646--1654.

\bibitem{bib15}
F.~Yang, H.~Yang, J.~Fu, H.~Lu, and B.~Guo, ``Learning texture transformer network for image super-resolution,'' in \emph{Proceedings of the IEEE/CVF conference on computer vision and pattern recognition}, 2020, pp. 5791--5800.

\bibitem{bib32}
H.~Zheng, M.~Ji, H.~Wang, Y.~Liu, and L.~Fang, ``Crossnet: An end-to-end reference-based super resolution network using cross-scale warping,'' in \emph{Proceedings of the European conference on computer vision (ECCV)}, 2018, pp. 88--104.

\bibitem{bib33}
Z.~Zhang, Z.~Wang, Z.~Lin, and H.~Qi, ``Image super-resolution by neural texture transfer,'' in \emph{Proceedings of the IEEE/CVF conference on computer vision and pattern recognition}, 2019, pp. 7982--7991.

\bibitem{bib34}
L.~Lu, W.~Li, X.~Tao, J.~Lu, and J.~Jia, ``Masa-sr: Matching acceleration and spatial adaptation for reference-based image super-resolution,'' in \emph{Proceedings of the IEEE/CVF Conference on Computer Vision and Pattern Recognition}, 2021, pp. 6368--6377.

\bibitem{bib14}
A.~Vaswani, N.~Shazeer, N.~Parmar, J.~Uszkoreit, L.~Jones, A.~N. Gomez, {\L}.~Kaiser, and I.~Polosukhin, ``Attention is all you need,'' \emph{Advances in neural information processing systems}, vol.~30, 2017.

\bibitem{bib43}
F.~Jia, L.~Tan, G.~Wang, C.~Jia, and Z.~Chen, ``A super-resolution network using channel attention retention for pathology images,'' \emph{PeerJ Computer Science}, vol.~9, p. e1196, 2023.

\bibitem{bib7}
I.~Goodfellow, J.~Pouget-Abadie, M.~Mirza, B.~Xu, D.~Warde-Farley, S.~Ozair, A.~Courville, and Y.~Bengio, ``Generative adversarial nets,'' \emph{Advances in neural information processing systems}, vol.~27, 2014.

\bibitem{bib25}
A.~Hore and D.~Ziou, ``Image quality metrics: Psnr vs. ssim,'' in \emph{2010 20th international conference on pattern recognition}.\hskip 1em plus 0.5em minus 0.4em\relax IEEE, 2010, pp. 2366--2369.

\bibitem{bib26}
Z.~Wang, A.~C. Bovik, H.~R. Sheikh, and E.~P. Simoncelli, ``Image quality assessment: from error visibility to structural similarity,'' \emph{IEEE transactions on image processing}, vol.~13, no.~4, pp. 600--612, 2004.

\bibitem{bib18}
Z.~Liu, Y.~Lin, Y.~Cao, H.~Hu, Y.~Wei, Z.~Zhang, S.~Lin, and B.~Guo, ``Swin transformer: Hierarchical vision transformer using shifted windows,'' in \emph{Proceedings of the IEEE/CVF international conference on computer vision}, 2021, pp. 10\,012--10\,022.

\bibitem{bib19}
J.~Cao, Y.~Li, K.~Zhang, and L.~V. Gool, ``Video super-resolution transformer,'' \emph{CoRR}, vol. abs/2106.06847, 2021.

\bibitem{bib35}
M.~J. Shensa \emph{et~al.}, ``The discrete wavelet transform: wedding the a trous and mallat algorithms,'' \emph{IEEE Transactions on signal processing}, vol.~40, no.~10, pp. 2464--2482, 1992.

\bibitem{bib36}
V.~Nair and G.~E. Hinton, ``Rectified linear units improve restricted boltzmann machines,'' in \emph{Proceedings of the 27th international conference on machine learning (ICML-10)}, 2010, pp. 807--814.

\bibitem{bib37}
Y.~Blau, R.~Mechrez, R.~Timofte, T.~Michaeli, and L.~Zelnik-Manor, ``The 2018 pirm challenge on perceptual image super-resolution,'' in \emph{Proceedings of the European Conference on Computer Vision (ECCV) Workshops}, 2018, pp. 0--0.

\bibitem{bib38}
G.~Litjens, P.~Bandi, B.~Ehteshami~Bejnordi, O.~Geessink, M.~Balkenhol, P.~Bult, A.~Halilovic, M.~Hermsen, R.~van~de Loo, R.~Vogels \emph{et~al.}, ``1399 h\&e-stained sentinel lymph node sections of breast cancer patients: the camelyon dataset,'' \emph{GigaScience}, vol.~7, no.~6, p. giy065, 2018.

\bibitem{bib39}
K.~Sun, Y.~Gao, T.~Xie, X.~Wang, Q.~Yang, L.~Chen, K.~Wang, and G.~Yu, ``A low-cost pathological image digitalization method based on 5 times magnification scanning,'' \emph{Quantitative Imaging in Medicine and Surgery}, vol.~12, no.~5, p. 2813, 2022.

\bibitem{bib40}
B.~S. Veeling, J.~Linmans, J.~Winkens, T.~Cohen, and M.~Welling, ``Rotation equivariant cnns for digital pathology,'' in \emph{Medical Image Computing and Computer Assisted Intervention--MICCAI 2018: 21st International Conference, Granada, Spain, September 16-20, 2018, Proceedings, Part II 11}.\hskip 1em plus 0.5em minus 0.4em\relax Springer, 2018, pp. 210--218.

\bibitem{bib41}
R.~Zhang, P.~Isola, A.~A. Efros, E.~Shechtman, and O.~Wang, ``The unreasonable effectiveness of deep features as a perceptual metric,'' in \emph{Proceedings of the IEEE conference on computer vision and pattern recognition}, 2018, pp. 586--595.

\end{thebibliography}

\begin{IEEEbiography}
[{\includegraphics[width=1in,height=1.25in,clip,keepaspectratio]{{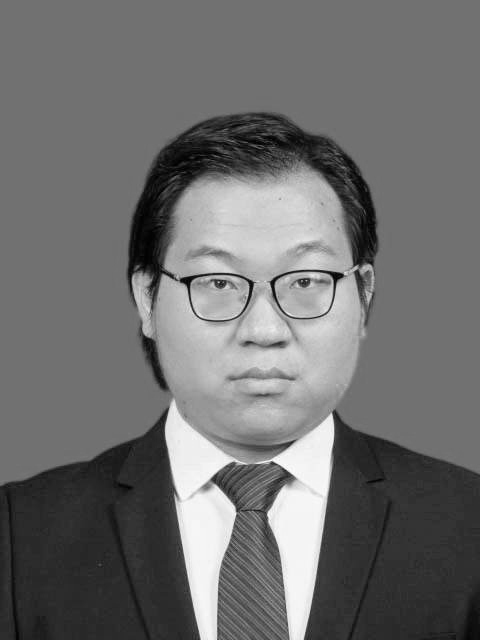}}}]
{Feiyang Jia} was born in Yinchuan, Ningxia Province, China, in 1998. He received his B.S. degree from Beijing Jiaotong University (China) in 2020. He received a master's degree from Beijing Technology and Business University (China) in 2023. He is now a Ph.D. student majoring in Computer Science and Technology at Beijing Jiaotong University (China), with research focus on Computer Vision.
\end{IEEEbiography}

\begin{IEEEbiography}[{\includegraphics[width=1in,height=1.25in,clip,keepaspectratio]{{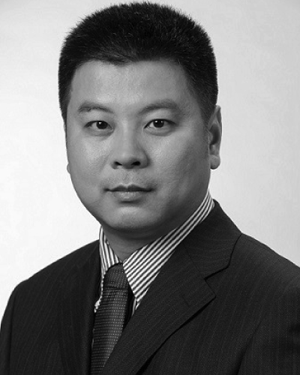}}}]{Zhineng Chen} (Member, IEEE) received the M.Sc. and B.Sc. degrees in computer science from the College of Information Engineering, Xiangtan University, Xiangtan, China, in 2004 and 2007, respectively, and the Ph.D. degree in computer science from the Institute of Computing Technology, Chinese Academy of Sciences, Beijing, China, in 2011. He was an Associate Professor with the Institute of Automation, Chinese Academy of Sciences, and was a Senior Research Associate with the Department of Computer Science, City University of Hong Kong, Hong Kong. He is currently a pretenured Professor with the School of Computer Science, Fudan University, Shanghai, China. He has published over 60 academic articles in prestigious journals and conferences. His research interests include large-scale multimedia analytics, medical image analysis, and pattern recognition.
\end{IEEEbiography}

\begin{IEEEbiography}[{\includegraphics[width=1in,height=1.25in,clip,keepaspectratio]{{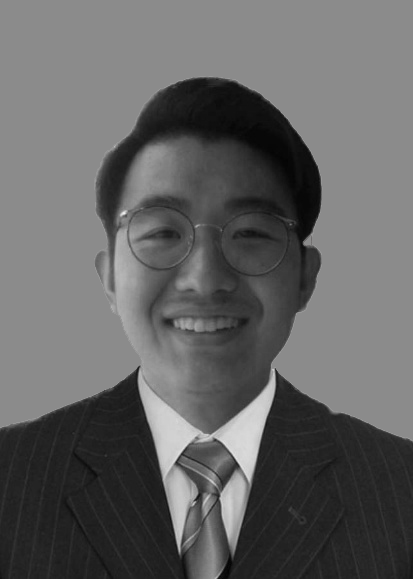}}}]{Ziying Song} was born in Xingtai, Hebei Province, China, in 1997. He received his B.S. degree from Hebei Normal University of Science and Technology (China) in 2019. He received a master's degree from Hebei University of Science and Technology (China) in 2022. He is now a Ph.D. student majoring in Computer Science and Technology at Beijing Jiaotong University (China), with research focus on Computer Vision. 
\end{IEEEbiography}

\begin{IEEEbiography}
[{\includegraphics[width=1in,height=1.25in,clip,keepaspectratio]{{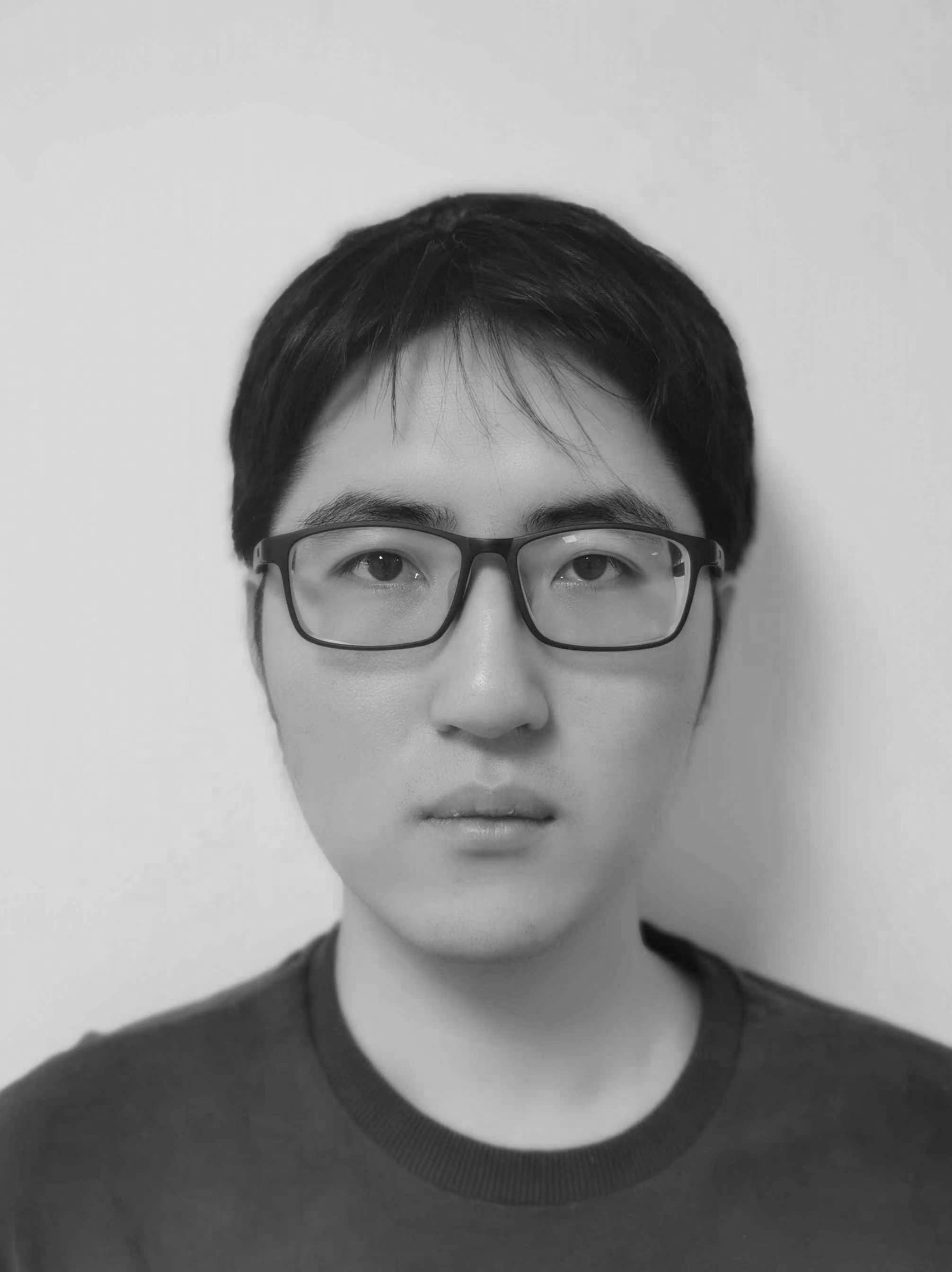}}}] {Lin Liu} was born in Jinzhou, Liaoning Province, China, in 2001. He is now a college student majoring in Computer Science and Technology at China University of Geosciences(Beijing). Since Dec. 2022, he has been recommended for a master's degree in Computer Science and Technology at Beijing Jiaotong University. His research interests are in computer vision.
\end{IEEEbiography}

\begin{IEEEbiography}[{\includegraphics[width=1in,height=1.25in,clip,keepaspectratio]{{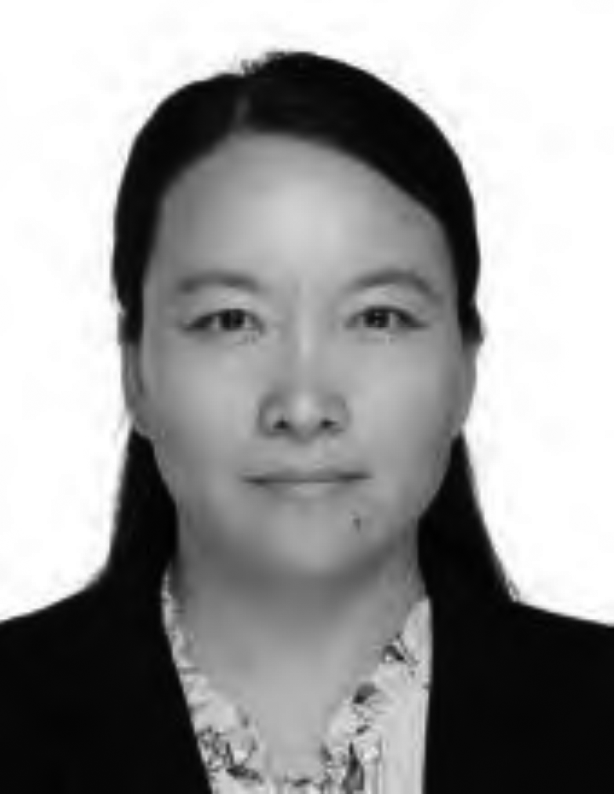}}}]{Caiyan Jia} was born in 1976. She received her Ph.D. degree from Institute of Computing Technology, Chinese Academy of Sciences, China, in 2004. She had been a postdoctor in Shanghai Key Lab of Intelligent Information Processing, Fudan University, Shanghai, China, in 2004–2007. She is now a professor in School of Computer and Information Technology, Beijing Jiaotong University, Beijing, China. Her current research interests include deep learning in computer vision, graph neural networks and social computing, etc.
\end{IEEEbiography}

\end{document}